\documentclass[intlimits,twoside,a4paper]{article}

\usepackage{amsmath,amssymb}
\usepackage{graphicx}

\usepackage[T2A]{fontenc}
\usepackage[cp1251]{inputenc}

\usepackage{cmpj2}

\issue{2013}{16}{1}{14001}

\doinumber{10.5488/CMP.16.14001}

\articletype{Rapid Communication}

\title[van der Waals loop]%
{A non-classical van der Waals loop: \\Collective variables method%
}
  \author{R.V. Romanik, M.P. Kozlovskii}
  \address{Institute for Condensed Matter Physics of the National Academy of Sciences of Ukraine, \\
  1 Svientsitskii St., 79011 Lviv, Ukraine}

\date{Received December 7, 2012, in final form February 5, 2013}
\authorcopyright{R.V. Romanik, M.P. Kozlovskii, 2013}

\begin{document}

\newcommand{\sgn}{\mathop{\mathrm{sgn}}}

\maketitle

\begin{abstract}
The equation of state is investigated for an Ising-like model in the framework of collective variables method.
The peculiar feature of the theory is that a non-classical van der Waals loop is extracted.
The results  are compared with the ones of a trigonometric parametric model in terms
of normalized magnetization, $\tilde{M},$ and field,~$\tilde{H}.$
\pacs 05.50.+q, 05.70.Ce, 64.60.F-, 75.10.Hk
\end{abstract}

Recently, in~\cite{CMP_11} it was shown that considering the system of Ising spins in an external field within the collective variables (CV) method \cite{Yukh_87,Mic_Theory_01}, a contribution to free energy can be singled out that is an analogue of Landau free energy. As a consequence, a non-classical van der Waals (vdW) loop is obtained. In the present paper, the shape of the loop is investigated. A comparison is made with similar results of~\cite{Fish_Zinn_99}, where vdW loop was obtained for a system with a scalar order parameter using a trigonometric parametric model for scaling behaviour near criticality. The discussion on whether such a loop exists naturally and whether it has any physical manifestations can be found in~\cite{Fish_Zinn_98}. Our purpose in this paper is to compare the results from CV theory with the other ones.

We consider a system of $N$ Ising spins on a simple cubic lattice of spacing $c.$ The Hamiltonian of such a system is
well known
\begin{equation}
\label{H}
  H=-\frac12\sum_{{i},{j}}\Phi(r_{{i}{j}})\sigma_{{i}}\sigma_{{j}}
  -{\cal{H}}\sum_{{i}}\sigma_{{i}}\,.
\end{equation}
Here, the spin variables $\sigma_i$ take on $\pm 1,$ ${\cal{H}}$ is the external field, and
\(\Phi(r_{{i}{j}})\) is a short-range interaction potential between spins located at
the $i$-th and $j$-th sites of separation $r_{ij}.$ The interaction potential can be chosen in the form of exponentially decreasing function, $\Phi(r_{ij})=\mathrm{const}\cdot \exp{(-r_{ij}/b)},$ with $b$ being an effective range.

The partition function $Z=\sum_{\{\sigma\}}\rm e^{-\beta H},$
where $\beta=(k_{\rm B}T)^{-1}$ is the inverse temperature,
can be written in terms of collective variables $\rho_{\mathbf{k}}$ \cite{Yukh_87,Mic_Theory_01}. In ``$\rho^4$-model'' approximation, the explicit form for such a representation is as follows:
\begin{eqnarray}
\label{Z_func}
Z=Z_0\int({\rm d}\rho)^{N_0}\exp
\left[a_1\sqrt{N_0}\rho_0-\frac{1}{2}\sum_{\bf{k}\in{\cal{B}}_0}d(k)\rho_{\bf{k}}\rho_{-\bf{k}}
-\frac{a_4}{4!}N_0^{-1}\sum_{{\bf{k}}_i
\in{\cal{B}}_0}\rho_{{\bf{k}}_1}\ldots\rho_{{\bf{k}}_4}\delta_{{\bf{k}}_1+\ldots+{\bf{k}}_4}
\right].
\end{eqnarray}
Here, the quantity $d(k)$ contains the Fourier transform of the interaction potential
\begin{equation}
\label{d_k}
d(k)=a_2+\beta\tilde{\Phi}(0)\bar{\Phi}-\beta\tilde{\Phi}(k).
\end{equation}
For $\tilde{\Phi}(k)$, we use the so-called parabolic approximation
\begin{equation}
\label{Phi_par}
  \tilde{\Phi}(k)=\left\{\begin{array}{ll}
                      \tilde{\Phi}(0)\left(1-2b^2k^2\right), & k\leqslant{{B}}_0={{B}}s_0^{-1}, \\
                      \Phi_0=\tilde{\Phi}(0)\bar{\Phi}, & {{B}}_0<{k}\leqslant {{B}},
                      \end{array}
                   \right.
\end{equation}
where $\Phi_0$ is some average value for $\tilde{\Phi}(k)$ with large $k,$ which is defined by parameter $\bar{\Phi}.$

Strictly speaking, the wave vector $\bf{k}$ takes on the values from the first Brillouin zone
\begin{eqnarray}
\label{B_zone1}
  {\cal{B}}=\{{\bf{k}}=(k_x,k_y,k_z)| \ k_i=-\frac{\pi}{c}+\frac{2\pi}{c}\frac{n_i}{N_i}; \  n_i=1,2,\ldots,N_i; \ N_i^3=N; i=x,y,z\}.
\end{eqnarray}
In what follows, however, we will keep to the spherical approximation for the Brillouine zone so that $B=\pi/c$ is the boundary of this zone, and $B_0=\pi/(cs_0)=\pi/c_0$ is the boundary of ${\cal{B}}_0.$
The discussion on the choice for $s_0$ and $\bar{\Phi}$ can be found in \cite{CMP_10}. In general, $s_0$ should depend on the ratio of the effective interaction range $b$ to the lattice constant $c.$ In the present calculation, we fix $b/c=0.3$ and $s_0=2.$ This yields $\bar{\Phi}=0.329$ and numerical values for other coefficients needed to represent the results are presented in table~\ref{tab_coefs}. The quantities $a_i$ from~(\ref{Z_func}) are expressed as follows:
\begin{equation}
a_1=s_0^{d/2}h, \qquad a_2=1-s_0^{-d}, \qquad a_4=2s_0^{-d},
\end{equation}
where $d=3$ is the space dimension, and $h=\beta{\cal{H}}$ is dimensionless field. In~(\ref{Z_func}), the collective variables $\rho_{\bf{k}}$ with ${{B}}_0<k\leqslant{{B}}$ have already been integrated out so that $N_0=Ns_0^{-3}$ is a number of variables remaining to be integrated.

\begin{table}[!t]
\caption{Numerical values of some quantities used in the present calculations.}
\label{tab_coefs}
\begin{center}
\begin{tabular}{|c|c|c|c|c|c|c|c|c|c|c|c|c|c|}
\hline\hline $b/c$ & $s_0$ & $\beta_\mathrm{c}\Phi(0)$ & $\overline{\Phi}$ &  $s$ & $E_1$  &$E_2$ \\
\hline
0.3 & 2.0 & 1.6411 & 0.32898 & 3.5977 & 24.551 & 8.306 \\
\hline
\hline
$f_0$ & $h_0$ & $c_{1k}$ & $\varphi_0$ & $\Phi_f$ & $n_0$ & \\
\hline
0.5& 0.760 & 1.176 &0.5938 & 0.105 & 0.5 &\\
\hline\hline
\end{tabular}
\end{center}
\end{table}

Calculation of the partition function, equation~(\ref{Z_func}), is performed according to Yukhnovskii's method~\cite{Yukh_87}. It is based on the idea of step-by-step integration of the partition function over the subsets of collective variables, first with $B_1<k\leqslant B_0,$ then with $B_2<k\leqslant B_1,$ and so on while averaging the Fourier transform of the interaction potential on each step (a consequence of this averaging is that the critical exponent $\eta,$ characterizing the decay of correlation length, equals zero~--- as is in the case of local potential approximation). Here, $B_n=B_{n-1}/s=B_0/s^n,$ where $s$ is the renormalization group (RG) parameter. This is equivalent to the Kadanoff scheme of constructing spin blocks~\cite{Kad_66,Kad_67}. Every time when integrating over a subset of CV, a factor~--- let us denote it by $Q_n$~--- appears in the partition function.
On performing step-by-step integration of the partition function over $n_{\rm p}$ subsets, one arrives at
\begin{equation}
\label{Z3}
Z=Z_0\big[Q(d)\big]^{N_0}\left(\prod_{n=1}^{n_{\rm p}}Q_n\right)Z_{\rm LGR}\,,
\end{equation}
where $Z_0$ and $Q(d)$ give analytical contributions to free energy and are not important for the critical behavior, $Q_n$ is the partial partition function due to fluctuations with $B_n<k \leqslant B_{n-1},$ $Z_{\rm LGR}$ is the contribution from $k$ small~--- the limiting (inverse) Gaussian regime of fluctuations.
Each of the partial partition functions $Q_n$ is characterized by its own set of coefficients $a_1^{(n)},$ $d_n(0)$, $a_4^{(n)}$, for which the recurrence relations (RR) hold \cite{CMP_05}. The RR have a fixed point as a partial solution. That is why the quantity $n_{\rm p}$ is chosen from the requirement that for $n\leqslant n_{\rm p}$  the RR can be linearized near the fixed point. In this case, the system possesses the RG symmetry and is said to be in the critical regime of the order parameter fluctuations. The quantity $n_{\rm p}$ is called the exit point from the critical regime.

The analytic expressions for the exit point were found for the limiting cases, $h=0$~\cite{Yukh_87} and $\tau=0$~\cite{PhaseTrans_07}
\begin{equation}\label{nplim}
n_{\rm p}=\left\{
            \begin{array}{ll}
              -\dfrac{\ln{|\tilde{h}|}}{\ln{E_1}}-1, & \hbox{ {at} \quad $\tau=0$,} \\[2ex]
              -\dfrac{\ln{|\tilde{\tau}|}}{\ln{E_2}}-1, & \hbox{ {at} \quad $h=0$,}
            \end{array}
          \right.
\end{equation}
where $\tau=(T-T_{\rm c})/T_{\rm c}$ is the reduced temperature, $E_1=24.551$ and $E_2=8.306$ are the eigenvalues of the matrix of the RG transformation linearized near the fixed point of RR,
\begin{equation}
\tilde{\tau}=\frac{c_{1{\rm k}}}{f_0}\tau, \qquad \tilde{h}=\frac{s_0^{d/2}}{h_0}h,
\end{equation}
where $f_0=0.5$ defines the fixed-point coordinates, $c_{1{\rm k}}=1.176,$ $h_0=0.760.$
In general case, the expression for $n_{\rm p}$ cannot be obtained analytically. For example, in~\cite{NPB_06} this quantity was computed numerically as a solution to a certain equation. In any case, $n_{\rm p}$ should satisfy the conditions~(\ref{nplim}) in the mentioned limiting cases. Based on this requirement, in~\cite{PhaseTrans_07} the expression was constructed for the exit point in the form
\begin{equation}
\label{exit_p}
n_{\rm p}^{(\pm)}=-\frac{\ln{\left(\tilde{h}^2+{h^{(\pm)}_{\rm c}}^2\right)}}{2\ln{E_1}}-1,
\end{equation}
where some temperature fields, $h_{\rm c}^{(+)}=|\tau|^{\beta\delta}$ and $h_{\rm c}^{(-)}=|\tau_1|^{\beta\delta},$ are introduced, $\beta$ and $\delta$ are the critical exponents\footnote{Denotation $\beta$ for the temperature critical exponent of magnetization is widespread in literature, the same notation $\beta$ is also widely accepted for the inverse temperature. We hope that the context will prevent the reader from confusing one quantity with another.} of magnetization. In our approach,
\begin{equation}
\label{crit_exp}
\beta=0.302 \quad \text{and} \quad \delta=d+2=5,
\end{equation}
where $d=3$.
The signs ``$+$'' and ``$-$'' are related to $T>T_{\rm c}$ and $T<T_{\rm c},$ respectively. In what follows, we will mainly omit the superscript $\pm$. Finally,  $\tau_1=-E_2^{n_0}\tau,$ where $n_0$ denotes the difference between $n_{\rm p}^{(+)}$ and $n_{\rm p}^{(-)}$ for $h=0.$ In~\cite{UFZ_09}, $n_0=0.5$ is chosen to recover the universal ratio of critical amplitudes for the correlation length, $\xi^+/\xi^-=1.896(10)$~\cite{CPRV_02}.

In expression~(\ref{Z3}), the quantity $Z_{\rm LGR}$ is still to be expressed. A detailed explanation of how to compute it can be found in \cite{UFZ_09,CMP_09}. We just recall that it is expressed in the form
\begin{equation} \label{Z_LGR}
Z_{\rm LGR}=Z_{\rm G}Z_{\rm TR}\cdot \rm e^{NE_0(\sigma)},
\end{equation}
where $Z_{\rm TR}$ is from the so-called transition region of fluctuations, $Z_{\rm G}$ is from the region of $k$ small, and the contribution $\re^{NE_0(\sigma)}$ is the most important due to the collective variable $\rho_0.$ In the present research, main attention is paid to this part of the partition function. The quantity $E_0(\sigma)$ has the form
\begin{equation}
E_0=e_0h\left(\tilde{h}^2+{h^{(\pm)}_{\rm c}}^2\right)^{\frac{1}{2(d+2)}}-e_2\left(\tilde{h}^2+{h^{(\pm)}_{\rm c}}^2 \right)^{\frac{d}{d+2}}.
\end{equation}
The following notation is used
\begin{eqnarray}
e_0=\sigma_0s^{-1/2}, \quad \qquad
e_2=\frac{1}{2}\sigma_0^2s^{-3}\left(r_{n_{\rm p}+2}+\frac{1}{12}s_0^3\sigma_0^2u_{n_{\rm p}+2}\right),
\end{eqnarray}
where
\begin{eqnarray}
r_{n_{\rm p}+2}&=&\beta\tilde{\Phi}(0)f_0\left(-1\pm E_2 H_{\rm c}\right), \nonumber \\
u_{n_{\rm p}+2}&=&\left[\beta\tilde{\Phi}(0)\right]^2\varphi_0\left(1\pm \Phi_f E_2 H_{\rm c}\right)
\end{eqnarray}
with $H_{\rm c}=\tilde{\tau}\left(\tilde{h}^2+{h^{(\pm)}_{\rm c}}^2\right)^{-1/2\beta\delta}.$ The quantity $\sigma_0$ is the solution of the following cubic equation
\begin{equation}
\label{cubic}
\sigma_0^3+p\sigma_0+q=0
\end{equation}
with the coefficients
\begin{eqnarray}
p=6s_0^{-3}\frac{r_{n_{\rm p}+2}}{u_{n_{\rm p}+2}}\,, \qquad
q=-6s_0^{-9/2}s^{5/2}\frac{h_0}{u_{n_{\rm p}+2}}\frac{\tilde{h}}{\left(\tilde{h}^2+{h^{(\pm)}_{\rm c}}^2\right)^{1/2}}\,.
\end{eqnarray}

Based on~(\ref{Z3})~and~(\ref{Z_LGR}), the Gibbs free energy can be expressed as a sum of three terms
\begin{equation}
\label{free_en}
F(\tau,h)=-k_{\rm B}T\ln{Z}=F_{\rm a}+F_{\rm s}^{(\pm)}+F_0^{(\pm)}.
\end{equation}
Here, the term $F_{\rm a}$ is the analytical part of the free energy and does not affect the critical behaviour. The term $F_{\rm s}^{(\pm)}$ is expressed as
\begin{equation}
\label{Fs}
F_{\rm s}^{(\pm)}(\tau,h)=-k_{\rm B}TN\gamma_{\rm s}^{(\pm)}\left(\tilde{h}^2+{h^{(\pm)}_{\rm c}}^2\right)^{\frac{d}{d+2}},
\end{equation}
where $\gamma_{\rm s}^{\pm}$ includes contributions from the critical regime and from the limiting Gaussian regime (inverse Gaussian regime in the case of $T<T_{\rm c}$). The explicit expression for it can be found in~(5.6) of~\cite{CMP_09}.
Finally, the quantity $F_0^{(\pm)}$ from~(\ref{free_en}) is
\begin{equation}
\label{F0}
F_0^{(\pm)}(\tau,h)=-k_{\rm B}TNE_0(\sigma).
\end{equation}
This contribution to the Gibbs free energy is due to the collective variable $\rho_0,$ which, as is known from the theory of collective variables~\cite{Yukh_87}, is related to the order parameter. Therefore, $F_0^{(\pm)}$ is the free energy of ordering and can be regarded as the analogue of Landau free energy. In the case of zero external field, such an analogue was found earlier in~\cite{PRB_02}. The order parameter of the considered system~--- the magnetization~--- is then calculated by means of the thermodynamic formula
\[
M_0=-\frac{1}{N}\frac{\partial{F_0^{(\pm)}}}{\partial{{\cal{H}}}}=\frac{\partial{E_0(\sigma)}}{\partial{h}}\,,
\]
which leads to the expression that can be written in a compact form as
\begin{equation}\label{M0}
M_0=\sigma_{00}\left(\tilde{h}^2+{h^{(\pm)}_{\rm c}}^2\right)^{\frac{1}{2\delta}},
\end{equation}
where $\delta=5$ is the critical exponent describing the field dependence of magnetization and
\begin{eqnarray}
\sigma_{00}=\left[e_0\left(1+\frac{1}{5}\frac{\tilde{h}^2}{\tilde{h}^2+{h^{(\pm)}_{\rm c}}^2}\right)-
\frac{6}{5}e_2\frac{s_0^{3/2}}{h_0}\frac{\tilde{h}}{\left(\tilde{h}^2+{h^{(\pm)}_{\rm c}}^2\right)^{1/2}}-\left(\frac{\partial e_2}{\partial h}\right)_{\sigma}\left(\tilde{h}^2+{h^{(\pm)}_{\rm c}}^2\right)^{1/2}\right].
\end{eqnarray}


\begin{figure}[!b]
\centerline{
\includegraphics[width=0.45\textwidth,angle=0]{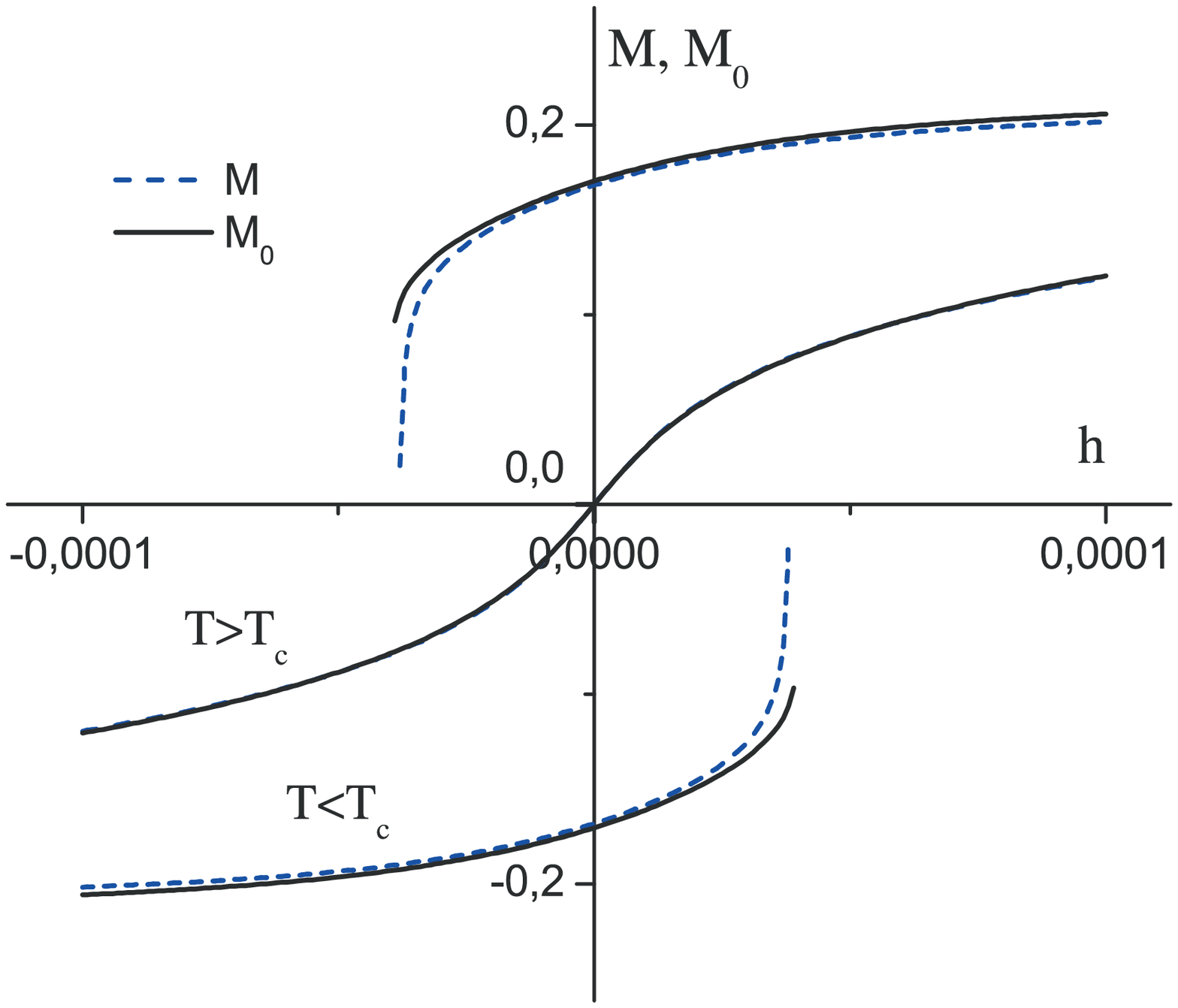} \hspace{0.05\textwidth}
\includegraphics[width=0.45\textwidth,angle=0]{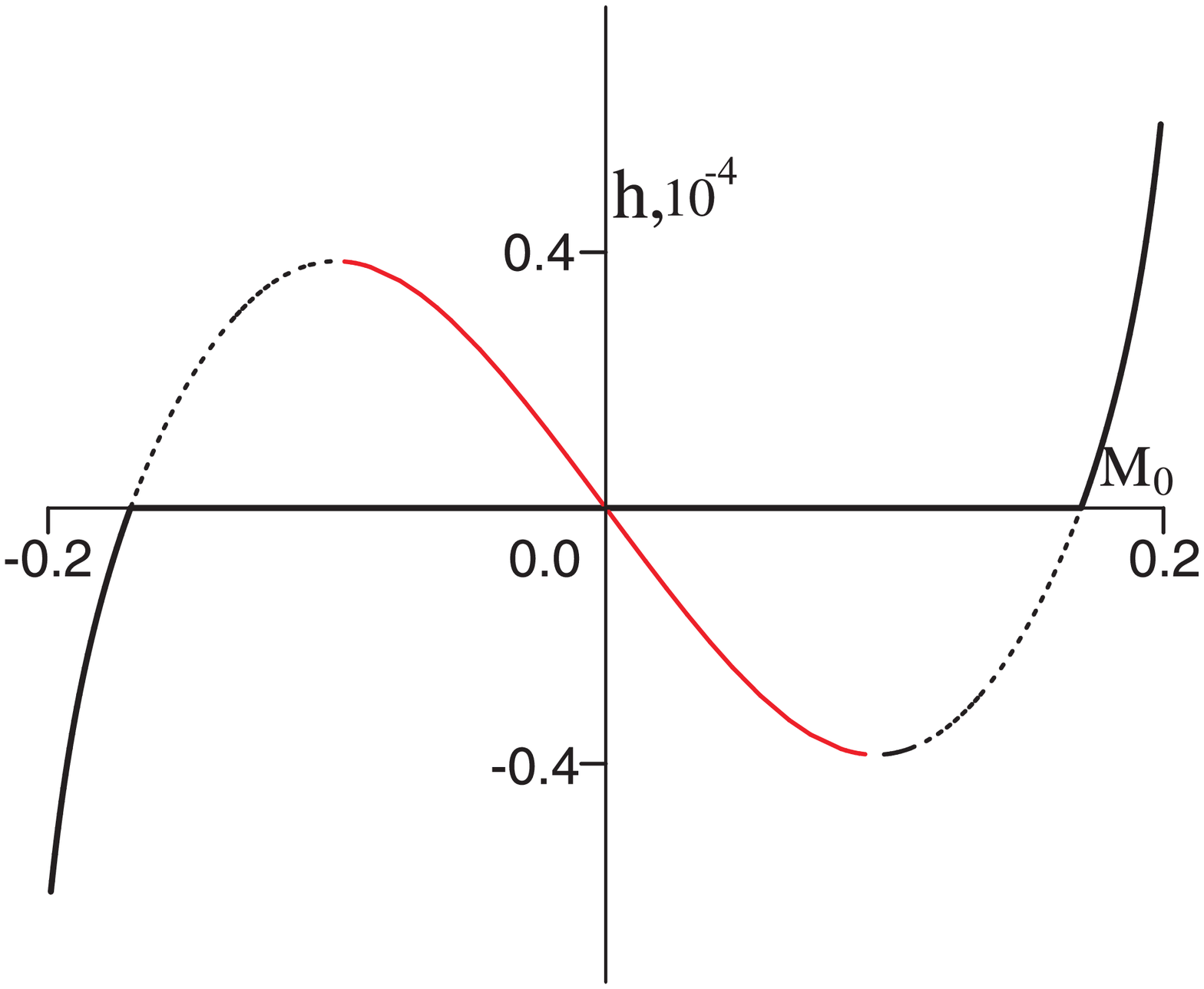} \\
}
\parbox[t]{0.51\textwidth}
{\caption{\label{M_full} (Color online) Dependence of magnetization on the field. Dashed line: full magnetization $M$, equation~(\ref{full_magn}); Solid line: contribution $M_0$ to magnetization due to Landau free energy, equation~(\ref{M0}). Temperatures are $\tau=\pm 0.001$.}} \hfill
\parbox[t]{0.48\textwidth}{\caption{\label{Mh} (Color online) Magnetization~(\ref{M0}) as  a function of the field below the critical temperature, $\tau=-0.001.$
}}
\end{figure}

We have compared $M_0$ with full magnetization $M$ defined by
\begin{equation}
\label{full_magn}
M=-\frac{1}{N}\frac{\partial F}{\partial {\cal{H}}}=-\frac{1}{N}\left(\frac{\partial F_\mathrm{s}^{(\pm)}}{\partial {\cal{H}}}+\frac{\partial F_0^{(\pm)}}{\partial {\cal{H}}}\right)  = M_\mathrm{s} + M_0\,.
\end{equation}
The results are demonstrated in figure~\ref{M_full}. As is seen, the magnetization of the system in stable states [magnetization and field have the same sign, $\sgn{(M)}=\sgn{(h)}$] is well described with $M_0$ alone. For $T>T_\mathrm{c}$, the difference between $M$ and $M_0$ is hardly observable in the scale of picture. For $T<T_\mathrm{c}$, the deviation of $M_0$ from $M$ is also minor in the regions of both $M>0$, $h>0$ and $M<0$, $h<0$. When the system goes into metastable states [$\sgn{(M)}=-\sgn{(h)}$], the situation becomes worse as the spinodal curve is approached, i.e., the curve of maximum magnitudes of the field for which a sign of magnetization can still be opposite to a sign of the field. In this case, the term $M_\mathrm{s}$ in~(\ref{full_magn}) becomes dominant.
This is due to the fact that we have used $Z_\mathrm{G}$ in the form as it was calculated in~\cite{CMP_09,JPS_09,UFZ_09}, with Gaussian measure. It is clear now that this accuracy is not sufficient to correctly account for the contributions from $F_\mathrm{s}^{(\pm)}$ in the metastable region. However, it is indeed sufficient in the stable region. Therefore, here we will present only the results of investigation on the basis of $F_0^{(\pm)}.$ Note, that in order to get the vdW loop, it is necessary to take into account all solutions of equation~(\ref{cubic}), but not only those minimizing the free energy.

%

In figure~\ref{Mh}, the magnetization $M_0,$ equation~(\ref{M0}), is presented as a function of the external field. As is seen from the picture, the proposed approach gives a van der Waals loop. However, the critical behaviour is characterized by non-classical critical exponents from equation~(\ref{crit_exp}).
In literature, arguments can be found that a non-classical theory of critical phenomena cannot give vdW loop~\cite{Isakov84} because there is no appropriate analytical continuation into the two-phase region. However, some phenomenological approaches have been suggested~\cite{Fish_Zinn_98,Fish_Zinn_99,Wycz_Anis_04} that incorporate both vdW loop and non-classical critical exponents.
We, in turn, have presented the microscopic approach in the framework of which a non-classical vdW loop can be obtained.

\begin{figure}[!b]
\vspace{-5mm}
\centerline{\includegraphics[width=0.62\textwidth,angle=0]{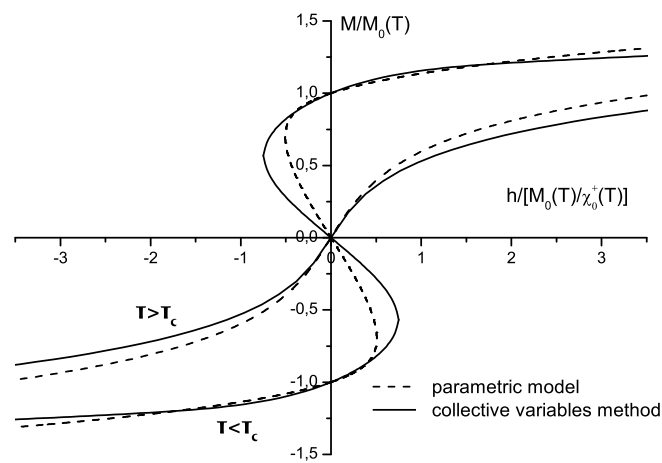}}
\caption{\label{loop} Scaled van der Waals loop ($T<T_\mathrm{c}$) and magnetization above the critical temperature ($T>T_\mathrm{c}$). The dashed curves represent the results obtained within the extended sine model~\cite{Fish_Zinn_99}. The full curves represent the results of the collective variables method, equation~(\ref{M0}).
 }
\end{figure}

In order to compare our results with the ones obtained by different methods, we appeal to work~\cite{Fish_Zinn_99}, where the loop was obtained with the help of a trigonometric parametric model for the scaling behaviour near criticality.
Figure~\ref{loop} presents the comparison. The dashed line is the result of Fisher with coworkers~\cite{Fish_Zinn_99}.
The solid line denotes our results based on equation~(\ref{M0}). Regarding the approximations, "$\rho^4$-model",
and zero value of the small critical exponent $\eta=0,$ we see a good qualitative agreement in the results, especially in the regions of $\tilde{h}\approx0$ and of $\tilde{h}\gg h_{\rm c}$, $\tilde{H}>2$. Some qualitative discrepancy is observed for the intermediate values of $\tilde{H}$, i.e., for $\tilde{h}\approx h_{\rm c}.$ In this domain, with increasing $\tilde{H}$, our curve initially withdraws from the dashed curve, further approaches and intersects it and then moves along with the dashed curve below it all the time. Such a discrepancy is connected with the choice of the functional form of $n_{\rm p},$ equation~(\ref{exit_p}). Although it provides correct values in the limiting cases, equation~(\ref{nplim}), it seems to fail in the intermediate region, $\tilde{h}\approx h_{\rm c}.$ Therefore, to improve our results we need a somewhat different functional form in this value range. Another way to do so is to compute $n_{\rm p}$ numerically, but we do not lose hope to solve the problem analytically and will attempt to find a more appropriate expression for the exit point in a future work.

Furthermore, we observe that the loop is wider in CV theory. The spinodal value of magnetization in CV, $\tilde{M}=0.55$ is less than the corresponding value $\tilde{M}=0.697$ from~\cite{Fish_Zinn_98}, which means that the CV theory, at least in "$\rho^4$-model" approximation, provides a wider metastable domain.

In conclusion, this work is the first attempt to investigate the van der Waals loop using the collective variables method, which is essentially a microscopic approach. This investigation is important because  there is lack of non-classical theories that give the vdW loop. Some quantitative disagreement of our result in comparison with the ones obtained in~\cite{Fish_Zinn_99}, can be associated with the model approximation, since the calculations are carried out using the simplest non-Gaussian approximation, i.e., $\rho^4$-model. The obtained result can be improved both in a formal way, by appropriately choosing the functional form of the exit point $n_{\rm p},$ and in a conceptual way, to which can be attributed (a) investigation of the system in a higher approximation, i.e., $\rho^n$-model  with $n>4$ (for $n=6$ see \cite{ISRN_11}), (b) the inclusion of corrections to scaling~\cite{PRB_02}, and (c) the averaging of the interaction potential~\cite{UFZ_12e}, the latter resulting in $\eta\neq 0.$ All of these will be the scope of a forthcoming paper.




\ukrainianpart

\title{Некласична петля ван дер Ваальса: \\ метод колективних змінних}
  \author{Р.В. Романік, М.П. Козловський}
  \address{Інститут фізики конденсованих систем НАН України, вул. І. Свєнціцького, 1,
79011 Львів, Україна}

\makeukrtitle

\begin{abstract}
\tolerance=3000%
Проводиться дослідження рівняння стану ізинґоподібної моделі в рамках методу колективних змінних. У використаному підході одержується  некласична форма петлі ван дер Ваальса. Проводиться порівняння, в термінах перенормованих змінних намагніченості і поля, отриманих результатів для петлі ван дер Ваальса з результатами тригонометричної параметричної моделі представлення критичної поведінки.

\end{abstract}

\end{document}